\pdfoutput=1
\documentclass[11pt]{article}
\DeclareUnicodeCharacter{2212}{-}
 
\usepackage{amsmath,amssymb}
\usepackage[top=20truemm,bottom=20truemm,left=20truemm,right=20truemm]{geometry}
\usepackage[pdftex]{graphicx}
\usepackage{amsmath}
\usepackage{here}
\usepackage{siunitx}
\usepackage[colorlinks=true
  ,urlcolor=blue
  ,anchorcolor=blue
  ,citecolor=blue
  ,filecolor=blue
  ,linkcolor=blue
  ,menucolor=blue
  ,linktocpage=true
]{hyperref}
\usepackage{cleveref}
\usepackage{subfigure}

\usepackage{color}
\usepackage{caption}
\usepackage[samesize]{cancel}
\usepackage{bm}
\usepackage{tikz-cd}
\usepackage{comment}
\usepackage{physics}
\usepackage{colortbl}
\usepackage{cite}

\definecolor{lightyellow}{rgb}{1.0, 0.95, 0.7}
\definecolor{lightblue}{rgb}{0.7, 0.9, 1.0}
\definecolor{lightpink}{rgb}{1.0, 0.85, 0.95}
\definecolor{lightgreen}{rgb}{0.7, 1.0, 0.4}
\definecolor{blue}{rgb}{0.0, 0.4, 1.0}
\definecolor{Blue}{rgb}{0,0,1}
\definecolor{darkgreen}{rgb}{0.,0.6,0.}

\usepackage{xcolor}
\usepackage{spectralsequences}
\usepackage{authblk}
\numberwithin{equation}{section}

\begin{document}

\title{Wess-Zumino-Witten Terms of $Sp$ QCD by Bordism Theory}
\author{Shota Saito}
\affil{Kavli Institute for the Physics and Mathematics of the Universe (WPI),\\The University of Tokyo,  Kashiwa, Chiba 277-8583, Japan}
\date{\today}
\maketitle

\begin{abstract}
    We investigate the four-dimensional Wess-Zumino-Witten (WZW) terms within the framework of $Sp$ quantum chromodynamics (QCD) using invertible field theory through bordism theory. We present a novel approach aimed at circumventing both perturbative and non-perturbative gauge anomalies on spacetime manifolds endowed with spin structures. We study both ungauged and gauged WZW terms including the problems of the topological consistency of gauged WZW terms.
\end{abstract}

\section{Introduction}

Quantum Chromodynamics (QCD) is a fundamental theory that elucidates the interactions among gauge bosons and charged fermions within a specific gauge group framework. Notably, in regimes with a sufficiently small number of flavors, QCD demonstrates asymptotic freedom, rendering it weakly coupled and amenable to analysis at high energies. However, as energy decreases to lower scales, QCD enters a regime of strong coupling, posing significant challenges to analysis. At these low energies, the theory undergoes notable changes in its spectrum, often described by a non-linear sigma model addressing Nambu-Goldstone scalar fields resulting from the spontaneous breaking of a flavor symmetry. Crucially, this flavor symmetry is chiral and subject to anomalies, which necessitate accurate reproduction at low energies following the anomaly matching conditions\cite{tH80}. When Nambu-Goldstone bosons comprise the sole massless spectrum, the description of anomalies is facilitated by the inclusion of the Wess-Zumino-Witten (WZW) term\cite{WZ71,Wit83a}. This term possesses topological characteristics, meaning it holds even when the background gauge fields of the flavor symmetry are turned off.

In recent years, an invertible QFT has attracted attention in condensed matter physics. This theory is characterized by a $U(1)$-valued partition function\cite{FM04}, typically observed within gapped ground states without degeneracy. Invertible topological phases which relies on background gauge fields are equivalent to symmetry protected topological (SPT) phases. In particular, when the target space corresponds to the classifying space $BG$ of group $G$, it is consistent with $G$-symmetry protected topological phases ($G$-SPT). In addition, the Wess-Zumino-Witten (WZW) term we are interested in serves as an example of invertible phases associated with non-linear sigma models. This is an invertible phase that depends on a background scalar field, representing a map from the spacetime to the target space of the sigma model. Including the above examples, the classification of invertible phases has been approached through the lens of bordism theory, a type of generalized cohomology theories\cite{KTTW14,FH16,GJ17,Yon18}. Notably, the WZW term in the non-linear sigma model is theorized to correspond to the anomaly. Anomalies associated with invertible phases have recently begun to be studied\cite{FKS17,Tho17,CFLS19a,CFLS19b,HKT20}.

The topological significance of the WZW term was initially highlighted by Witten, yet its applicability to arbitrary spacetime manifolds remained uncertain as Witten primarily focused on sigma models defined on spheres. It was subsequently revealed that to rigorously define the ungauged WZW term for $SU$ QCD and $SO$ QCD, the spacetime manifolds must possess spin structures\cite{Fre06,LOT20}. Here, we offer a precise formulation of the WZW term within the framework of $Sp$ QCD.

Investigating $Sp$ gauge theories plays a significant role in the development of particle phenomenology. This includes theoretical considerations regarding composite Higgs\cite{GK84,KG83,DGK84,Barnard:2013zea,Ferretti:2013kya}, top compositeness\cite{Kaplan:1991dc}, finite temperature phase transitions\cite{Holland:2003kg}, and so on. One reason for extending the Standard Model is the lack of explanation for the origin of dark matter\cite{Hambye:2008bq,Feng:2009mn,Cohen:2010kn,Foot:2014uba}. Dark matter can be described by a strongly coupled theory like QCD consisting only of SM singlets, potentially explained by composite dark matter or strongly interacting dark matter. Thus, studying $Sp$ gauge theories is also useful for dark matter research\cite{ZKMMNP22,DMZ23,Zie23}. Furthermore, the dark $Sp$ sector may undergo a phase transition due to dark confinement in the early universe, potentially leading to detectable background gravitational waves\cite{Kamionkowski:1993fg,Allen:1996vm,Schwaller:2015tja,Croon:2018erz,Christensen:2018iqi}. Over the past two decades, research on $Sp$ lattice gauge theories has been actively pursued\cite{Bennett:2017kga,Bennett:2019cxd,Bennett:2019jzz,Bennett:2020hqd,Bennett:2020qtj,Bennett:2021mbw,Bennett:2022ftz,Bennett:2022gdz,Bennett:2022yfa,Holland:2003kg,Hsiao:2022gju,Hsiao:2022kxf,Kulkarni:2022bvh,Lee:2018ztv,Lucini:2021xke,Maas:2021gbf}. These theories exhibit symmetry-breaking patterns different from those of $SU$ gauge theories, making their phenomenology highly intriguing. $Sp$ lattice gauge theories are currently under investigation, particularly regarding confinement/deconfinement phase transitions at finite temperature.

We adress in this paper the question whether the WZW term is accurately defined in $Sp$ gauge theories. It has been pointed out that the WZW term can be considered if an even number of flavors in the $Sp$ gauge theory undergo symmetry breaking in the most attractive channel\cite{Wit83b}. However, this has only been explored through considerations of homotopy groups, necessitating a more modern and precise analysis using cobordism theory. In this paper, we perform this precise analysis to verify the consistency of the low-energy effective theory of $Sp$ gauge theories. Additionally, this complements the results of the WZW term in the cases of $SU$ QCD and $SO$ QCD\cite{Fre06,LOT20}.

The structure of this paper is as follows. Note that our completely new studies are in Sec. \ref{sec4}. In Sec. \ref{sec2}, we explain the modern formulation of the WZW terms in invertible phases. It will be explained that the invertible phases are appropriately described by the Anderson dual of bordism groups, rather than by the conventional cohomological description. In Sec. \ref{sec3}, we review the case of $SU$ QCD as given by \cite{Fre06,LOT20}. We will see that the WZW terms are described by the invertible phases, and their definition requires the spin structures. In Sec. \ref{sec4}, we will discuss the WZW terms in $Sp$ QCD. We will see that the WZW terms are well-defined on the spin manifolds. In Appendix \ref{secA}, we collect the information of cohomology rings used in this paper, including classical Lie groups, classifying spaces, and homogeneous spaces. Much of this appendix is helped by \cite{MT91}. In Appendices \ref{secB} and \ref{secC}, we compute the required bordism group by Atiyah-Hirzebruch spectral sequence and Adams spectral sequence, respectively.

\section{Ungauged WZW terms as invertible phases and bordism theory}\label{sec2}

In this section, we first discuss the WZW terms based on homology theory. Recent research has shown that the classification of the WZW terms based on bordism theory yields correct results, rather than homology theory\cite{Fre06,LOT20}. Therefore, we redefine the WZW terms based on bordism theory. Furthermore, we demonstrate that the WZW terms can be expressed in terms of invertible phases. Utilizing bordism theory enables us to explore the structure of spacetime manifolds. In this paper, we focus on analyzing the spin bordism because we always assume a spin structure on a spacetime manifold. Please refer to the Appendices \ref{secB} and \ref{secC} for the computation method of spin bordism groups.

\subsection{The WZW terms via homology theory}\label{secWZWh}

Before transitioning to bordism groups to describe the WZW terms, we outline the conventional approach employing homology groups. Consider a $d$-dimensional theory with a scalar field $\phi$ taking values in a manifold $X$. Let $A$ be a $d$-form gauge field and $F$ be its closed $(d+1)$-form field strength on $X$ exhibiting the following property. When the scalar field $\phi:M_d\rightarrow X$ can be extended to $\phi:W_{d+1}\rightarrow X$ with $\partial W_{d+1}=M_d$, the WZW term takes the form:
\begin{equation}
\label{FH}
    e^{-S[\phi,A]}=e^{i\int_{\phi(W_{d+1})}F}.
\end{equation}
To ensure that this definition remains independent of the expansion of $M_d$, it is required that:
\begin{equation}
\label{HC}
    \int_{[C]}F\in2\pi\mathbb{Z},
\end{equation}
where $[C]\in H_{d+1}(X;\mathbb{Z})$.

Next, let us consider the scenario where the gauge field $A$ is flat and the field strength vanishes. In this case, the WZW term becomes independent of the deformation of $\phi(M_d)$. Suppose two $d$-dimensional scalar fields $\phi_0:M_d\rightarrow X$ and $\phi_1:M_d'\rightarrow X$, along with a $(d+1)$-dimensional scalar field $\Phi:W_{d+1}\rightarrow X$ satisfying the relation $W_{d+1}=M_d\cup \bar{M_d'}$, where $\bar{M_d'}$ denotes the orientation reversal of $M_d'$. We then have
\begin{equation}
    \frac{e^{-S[\phi_0,A]}}{e^{-S[\phi_1,A]}}=e^{i\int_{\Phi(W_{d+1})}F}.
\end{equation}
As a consequence, when $F$ vanishes, the WZW term becomes independent of the deformation, and the coupling is given by
\begin{equation}
\label{TH}
    e^{-S[\phi,A]}=\chi([\phi(M_d)]),
\end{equation}
where in mathematical terms $\chi:H_d(X;\mathbb{Z})\rightarrow U(1)$ corresponds to cohomology.

The $d$-form gauge field is classified by the topological class of $[F/2\pi]\in H^{d+1}(X;\mathbb{Z})$, and it can be decomposed into (\ref{FH}) and (\ref{TH}). This can be observed through the universal coefficient theorem
\begin{equation}
    0\rightarrow \mathrm{Ext}_\mathbb{Z}(H_d(X;\mathbb{Z}),\mathbb{Z})\rightarrow H^{d+1}_\mathbb{Z}(X;\mathbb{Z})\rightarrow \mathrm{Hom}_\mathbb{Z}(H_{d+1}(X;\mathbb{Z}),\mathbb{Z})\rightarrow0.
\end{equation}
We will encounter similar behavior later in the context of invertible phases or the Anderson dual of bordism groups.

\subsection{The Anderson dual of bordism groups and invertible phases}

It is believed that spin structures of spacetime manifolds are necessary for precisely defining the WZW terms. In other words, the spin bordism is considered more appropriate for formulating the WZW terms than the ordinary homology. Here, let us revisit what we discussed in the previous section from the perspective of the spin bordism.

In the ordinary cohomology theory, there are cases where the condition (\ref{HC}) is not satisfied by the WZW terms. However, considering spin structures on spacetime manifolds resolves such issues, indicating the suitability of the spin bordism. Just as with the ordinary homology, defining the WZW terms in the spin bordism requires two elements corresponding to (\ref{FH}) and (\ref{TH}) in the ordinary homology.

Let us start by examining the WZW terms associated with the free part. Just like before, we have the expression for the WZW term:
\begin{equation}
    e^{-S[\phi]}=e^{i\int_{\phi(W_{d+1})}F},
\end{equation}
where $F$ represents a closed $(d+1)$-form field strength, and $\phi:M_d\rightarrow X$ is a scalar field extendable to $\phi:W_{d+1}\rightarrow X$ such that $\partial W_{d+1}=M_d$. Here, $F$ is not only assumed to include differential forms on $X$ but also the Pontryagin classes of $\phi(W_{d+1})$. For this term to be well-defined regardless of the extension, it is necessary that
\begin{equation}
    \int_{\phi(W_{d+1})}F=2\pi\mathbb{Z},
\end{equation}
where $W_{d+1}$ is closed. Thus, we can construct the homomorphism
\begin{equation}
\label{FB}
    \mathrm{Hom}_\mathbb{Z}(\Omega_{d+1}^\mathrm{spin}(X),\mathbb{Z}).
\end{equation}
Since $\mathbb{Q}[p_1,p_2,\ldots]=H^\ast(BSpin;\mathbb{Q})$ where $p_i$ is the Pntryagin class of $\phi(W_{d+1})$, we can see that $F$ includes the information of $p_i$ taking values in $H^{d+1}(BSpin\times X;\mathbb{Q})=\mathrm{Hom}_\mathbb{Z}(\Omega_{d+1}^\mathrm{spin}(X),\mathbb{Z})\otimes\mathbb{Q}$.

When the field strength $F$ vanishes, the gauge field can be continuously deformed to a flat one, and the deformation class is classified by the torsion part. The WZW term determines the map $\Omega_d^\mathrm{spin}(X)\rightarrow U(1):[\phi]\mapsto e^{-S[\phi]}$, and it belongs to
\begin{equation}
\label{TB}
    \mathrm{Hom}_\mathbb{Z}(\Omega_d^\mathrm{spin}(X)_\mathrm{torsion},\mathbb{Z})=\mathrm{Ext}_\mathbb{Z}(\Omega_d^\mathrm{spin}(X),\mathbb{Z}).
\end{equation}

Mathematically, $\Omega_d^\mathrm{spin}(X)$ represents a generalized homology theory, and we can construct the associated generalized cohomology theory $D\Omega^d_\mathrm{spin}(X)$ (known as cobordism theory) by taking the Anderson dual:
\begin{equation}
    0\rightarrow\mathrm{Ext}_\mathbb{Z}(\Omega_d^\mathrm{spin}(X),\mathbb{Z})\rightarrow D\Omega^{d+1}_\mathrm{spin}(X)\rightarrow\mathrm{Hom}_\mathbb{Z}(\Omega_{d+1}^\mathrm{spin}(X),\mathbb{Z})\rightarrow0.
\end{equation}
It is worth noting that this is a generalization of the universal coefficient theorem for the ordinary homology theory.

On the other hand, the invertible phase is obtained by combining (\ref{FB}) and (\ref{TB}). This is achieved by situating the invertible phase in the middle of the short exact sequence:
\begin{equation}
\label{InvSES}
    0\rightarrow\mathrm{Ext}_\mathbb{Z}(\Omega_d^\mathrm{spin}(X),\mathbb{Z})\rightarrow\mathrm{Inv}^d_\mathrm{spin}(X)\rightarrow\mathrm{Hom}_\mathbb{Z}(\Omega_{d+1}^\mathrm{spin}(X),\mathbb{Z})\rightarrow0.
\end{equation}
By the definition, the invertible phase and the cobordism group are related by:
\begin{equation}
    \mathrm{Inv}^d_\mathrm{spin}(X)=D\Omega^{d+1}_\mathrm{spin}(X).
\end{equation}

Using the ordinary homology as previously done is inadequate in spin QFT because it also considers non-spin manifolds. In contrast, the spin bordism provides the appropriate framework. The WZW terms are precisely defined as invertible phases. From this point onward in this paper, we will conduct analyses using bordism theory.

\section{A brief review of the four-dimensional $SU(N_c)$ QCD case}\label{sec3}

In this section, we briefly review the WZW terms for $SU$ QCD. This work was initially presented by \cite{Fre06} and later rewritten in the modern language using invertible phases by \cite{LOT20}. The conventional definition of the WZW terms using homology theory is not precise. Here, we only present accurate results using bordism theory. First, we review the ungauged WZW terms as topological terms. Then, we demonstrate that the gauged WZW terms can be constructed from anomaly matching conditions. The detailed derivation is carefully explained in the next section on $Sp$ QCD.

\subsection{Anomalies in UV}

Consider the $SU(N_c)$ gauge theory with matters in the fundamental $N_c$-dimensional representation:
\begin{equation}
    \mathcal{L}=\int_{M_4}\mathrm{d}^4x\left(\tr F_{\mu\nu}F^{\mu\nu}+\bar{\psi}(i\cancel{D}-m)\psi\right).
\end{equation}
When the masses of the matters vanish, it is necessary to include an equal number of left-handed Weyl fermions and right-handed Weyl fermions to prevent gauge anomalies. Consequently, the flavor symmetry becomes $SU(N_f)_L\times SU(N_f)_R$. In this scenario, the coupling constant runs as
\begin{equation}
    \frac{\mathrm{d}g}{\mathrm{d}\log\mu}=-\frac{g^3}{48\pi^2}(11N_c-2N_f)+\mathcal{O}(g^4),
\end{equation}
where $\mu$ represents the energy scale. If the right-hand side of the above equation is negative, the theory exhibits asymptotic freedom. At that point, since the strong coupling typically occurs at low energies, making analysis difficult, we introduce a non-linear sigma model.

\subsubsection{Gauge anomalies}

The distinction between the cases of $N_c\ge3$ and $N_c=2$ is notable. For $N_c\ge3$, perturbative gauge anomalies exist, but there are no global gauge anomalies. The perturbative gauge anomaly is evident in the bordism group $\Omega_6^\mathrm{spin}(BSU(N_c))=\mathbb{Z}$, where the generator corresponds to the anomaly polynomial of a chiral fermion in the fundamental representation. Now we consider an equal number of right-handed and left-handed fermions in order to avoid the anomaly. Conversely, $\Omega_5^\mathrm{spin}(BSU(N_c))=0$ indicates the absence of global gauge anomalies.

In the case of $N_c=2$, unlike $N_c\ge3$, there are no perturbative gauge anomalies, but there are global gauge anomalies. The bordism group $\Omega_6^\mathrm{spin}(BSU(2))=0$ indicates the lack of perturbative gauge anomalies. However, in this scenario, anomalies persist. The presence of an odd number of Weyl fermions leads to a global gauge anomaly known as Witten's $SU(2)$ anomaly. The bordism group $\Omega_5^\mathrm{spin}(BSU(2))=\mathbb{Z}_2$ signifies that all gauge transformations fall into two classes: trivial or non-trivial. Under non-trivial transformations in a theory with odd Weyl fermions, the path integral yields the opposite sign compared to trivial transformations, rendering such a theory nonsensical.

\subsubsection{Global anomalies}

Consider four-dimensional $SU$ QCD with a flavor symmetry of $SU(N_f)_L\times SU(N_f)_R$, where $N_f\ge3$. The fermions are charged under $SU(N_c)\times SU(N_f)_L\times SU(N_f)_R$. The left-handed and right-handed chiral fermions are in the representations $\bm{N_c}\otimes\bm{N_f}^{(L)}\otimes\bm{1}^{(R)}$ and $\bm{\bar{N_c}}\otimes\bm{1}^{(L)}\otimes\bm{N_f}^{(R)}$ respectively. The corresponding anomaly polynomials are given by $N_c\times\mathrm{ch}_3(\bm{N_f}^{(L)})+N_f\times\mathrm{ch}_3(\bm{N_c})$ and $-N_c\times\mathrm{ch}_3(\bm{N_f}^{(R)})+N_f\times\mathrm{ch}_3(\bm{\bar{N_c}})$ respectively. Consequently, the anomaly polynomial of this theory becomes
\begin{equation}
\label{SUAN}
    N_c\left(\mathrm{ch}_3(\bm{N_f}^{(L)})-\mathrm{ch}_3(\bm{N_f}^{(R)})\right)=\frac{N_c}{2}\left(c_3^{(L)}-c_3^{(R)}\right),
\end{equation}
where $c_3^{(L,R)}$ represent the third Chern classes, and the equality holds due to the absence of the first Chern classes. Since $\Omega_5^\mathrm{spin}(B(SU(N_c)\times SU(N_f)))=0$, the anomaly polynomial completely determines the anomaly.

\subsection{Anomalies in IR}
As seen in (\ref{SUAN}), $SU$ QCD exhibits a global anomaly in the high-energy region. This anomaly must be reproduced at low energies by the anomaly matching conditions. Assuming that the only massless spectrum at low energies consists of Nambu-Goldstone scalar fields associated by the flavor symmetry breaking, they must reproduce the anomaly. This is achieved through the WZW term. In this section, we discuss the ungauged WZW term, the gauged WZW term, and related topics.

\subsubsection{Ungauged WZW terms for $N_f\ge3$}

Let us explore the WZW term of four-dimensional $SU$ QCD. To accurately characterize this term on a general manifold, a spin structure is necessary.

Consider four-dimensional $SU(N_c)$ QCD with massless quarks of $N_f$ flavors. When $N_f$ is sufficiently small, at low energies, this theory is believed to undergo spontaneous breaking of the flavor symmetry and to be described by the non-linear sigma model with a target space of $SU(N_f)$.

Now, let us consider the case where $N_f\ge3$. Given a field configuration $\sigma:M_4\rightarrow SU(N_f)$ and an appropriate expansion $\sigma:W_5\rightarrow SU(N_f)$ satisfying $\partial W_5=M_4$, there exists a $SU(N_f)$-invariant five-form $\tr(\sigma^{-1}d\sigma)^5$, which is the generator of $H^5(SU(N_f);\mathbb{R})$. The WZW term with the correct coefficient is given by
\begin{equation}
    e^{-S[\sigma:M_4\rightarrow SU(N_f)]}:=\exp\left(2\pi i\cdot N_c\int_{W_5}\Gamma_5\right),
\end{equation}
where
\begin{equation}
\label{WZWgamma}
    \Gamma_{2n-1}:=\left(\frac{i}{2\pi}\right)^n\frac{(n-1)!}{(2n-1)!}\tr(\sigma^{-1}d\sigma)^{2n-1},
\end{equation}
normalized to integrate to one over the generator of $\pi_{2n-1}(SU(N_f))\simeq\mathbb{Z}$.

The normalization is defined by the generator of $\pi_{2n-1}(SU(N_f))$, not $H_5(SU(N_f);\mathbb{Z})$. Let us compute the WZW term using the generator of $H_5(SU(N_f);\mathbb{Z})$. The following map sends 1 to $(n-1)!$ times the generator:
\begin{equation}
    \pi_{2n-1}(SU(N_f))\rightarrow H_5(SU(N_f);\mathbb{Z}).
\end{equation}
Consequently, $\Gamma_5$ integrates to $1/2$ over the generator of $H_5(SU(N_f);\mathbb{Z})$. When $N_c$ is odd, this violates (\ref{HC}), which is based on homology. However, assuming that $W_5$ has the spin structure, $\Gamma_5$ integrates to an integer. This ensures that the WZW coupling for odd $N_c$ is well-defined. Therefore, we conclude that if we can find the extension $\sigma:W_5\rightarrow SU(N_f)$ for a given $\sigma:M_4\rightarrow SU(N_f)$, then the WZW term is well-defined.

Another important consideration is whether such an extension exists. By employing bordism theory, one can resolve this question. The reduced bordism group $\Tilde{\Omega}_4^\mathrm{spin}(SU(N_f))$ is trivial when $N_f\ge3$. Consequently, any configuration $\sigma:M_4\rightarrow SU(N_f)$ is bordant to a constant configuration $\sigma_0:M_4\rightarrow SU(N_f)$ that maps $M_4$ to a single point in $SU(N_f)$. This implies the existence of a five-dimensional manifold $W_5$ such that $\partial W_5=M_4\cup\bar{M_4}$. Now, the WZW term for the configuration $\sigma_0$ is trivial, and the WZW term for the non-trivial configuration $\sigma$ can be extended to $W_5$. Following this approach, one can construct the four-dimensional WZW term for $N_f\ge3$ on any arbitrary four-dimensional spin manifold.

\subsubsection{Ungauged WZW terms for $N_f=2$}

Now, let us examine four-dimensional $SU$ QCD in the case of $N_f=2$. Due to $\mathrm{dim}(SU(2))=\mathrm{dim}(S^3)=3$, there is no suitable $5$-form. However, there exists a non-trivial bordism group $\Tilde{\Omega}^\mathrm{spin}_4(SU(2))=\mathbb{Z}_2$, contributing to the discrete WZW term as follows.

The WZW term for $SU(2)$ QCD is expressed as
\begin{equation}
    e^{-S[\sigma:M_4\rightarrow SU(2)]}:=(-1)^{N_c[\sigma:M_4\rightarrow SU(2)]},
\end{equation}
where $[\sigma:M_4\rightarrow SU(2)]$ denotes the equivalence class in $\Tilde{\Omega}^\mathrm{spin}_4(SU(2))$. The mapping $\sigma:M_4\rightarrow SU(2)$ describes the skyrmions. Quantizing the skyrmions can be approached in two ways. One approach is to assign equal weight to all configurations of skyrmions, while the other is to weight them with a factor of $(-1)^F$. These two ways correspond to quantizing the skyrmions as bosons and fermions respectively. This choice introduces an additional discrete parameter necessary to fully define the path integral. It is important to note that, once again, the spin structure on the manifold $M_4$ is required to define the $SU(2)$ WZW term.

\subsubsection{Ungauged WZW terms as invertible phases}
We will understand the WZW terms in terms of invertible phases. We will again divide the discussion into the cases of $N_f\ge3$ and $N_f=2$. By substituting the required bordism group into the expression represented by (\ref{InvSES}), we obtain the WZW terms, as it is expressed in terms of invertible phases.

For $N_f\ge3$, the related short exact sequence is as follows:
\begin{equation}
    0\rightarrow\mathrm{Ext}_\mathbb{Z}(\Omega_4^\mathrm{spin}(SU(N_f)),\mathbb{Z})\rightarrow\mathrm{Inv}^4_\mathrm{spin}(SU(N_f))\rightarrow\mathrm{Hom}_\mathbb{Z}(\Omega_5^\mathrm{spin}(SU(N_f)),\mathbb{Z})\rightarrow0.
\end{equation}
Since $\mathrm{Ext}_\mathbb{Z}(\Omega_4^\mathrm{spin}(SU(N_f)),\mathbb{Z})=0$, $\mathrm{Inv}^4_\mathrm{spin}(SU(N_f))\cong\mathrm{Hom}_\mathbb{Z}(\Omega_5^\mathrm{spin}(SU(N_f)),\mathbb{Z})\cong\mathbb{Z}$. This result corresponds to the fact that the definition of the WZW term includes the integer $N_c$.

For $N_f=2$, the related short exact sequence is as follows:
\begin{equation}
    0\rightarrow\mathrm{Ext}_\mathbb{Z}(\Omega_4^\mathrm{spin}(SU(2)),\mathbb{Z})\rightarrow\mathrm{Inv}^4_\mathrm{spin}(SU(2))\rightarrow\mathrm{Hom}_\mathbb{Z}(\Omega_5^\mathrm{spin}(SU(2)),\mathbb{Z})\rightarrow0.
\end{equation}
Since $\mathrm{Hom}_\mathbb{Z}(\Omega_5^\mathrm{spin}(SU(2)),\mathbb{Z})=0$, $\mathrm{Inv}^4_\mathrm{spin}(SU(2))\cong\mathrm{Ext}_\mathbb{Z}(\Omega_4^\mathrm{spin}(SU(2)),\mathbb{Z})\cong\mathbb{Z}_2$. This result implies that in the previous section, we have the option to quantize the skyrmion either as a boson or as a fermion.

\subsubsection{Normalization of ungauged WZW terms}

To define the WZW terms, it is necessary to ensure the reproduction of the anomaly matching conditions. The relation between the WZW terms and the anomaly polynomials is understood from the perspective of transgressions in algebraic topology. That is, if the symmetry is spontaneously broken from $G$ to $H$, the WZW term $\Gamma\in H^{d+1}(G/H)$ is transgressed into the anomaly polynomial $\alpha\in H^{d+2}(BG)$ through transgression associated to the fibration: $G/H\rightarrow BH\rightarrow BG$.

Let us explore the normalization process of the ungauged WZW terms. We use the Leray-Serre spectral sequence (LSSS) for the fibration $G/H\rightarrow BH\rightarrow BG$. The $E_2$ page is given by $E_2^{p,q}=H^p(BG;H^q(G/H;\mathbb{Z}))$, which converges to $H^{p+q}(BH;\mathbb{Z})$. Fortunately, we have access to all the required cohomology rings, enabling the computation of the transgression $d_n:E_n^{0,n-1}\rightarrow E_n^{n,0}$.

In $SU$ QCD, let us examine the scenario where $G$ is $SU\times SU$ and $H$ is $SU$. In this case, the fibration takes the form $SU\rightarrow BSU\rightarrow B(SU\times SU)$. The LSSS is given by

\begin{equation}
	\begin{array}{ccc}
		E_2^{p,q}=H^p(B(SU\times SU);H^q(SU;\mathbb{Z})) & & H^{p+q}(BSU;\mathbb{Z})\\
		\begin{array}{c|ccccccc}
			6 & & & & & & & \\
			5 & \mathbb{Z} & & & & \mathbb{Z}^{\oplus2} & & \mathbb{Z}^{\oplus2} \\
			4 & & & & & & & \\
			3 & \mathbb{Z} & & & & \mathbb{Z}^{\oplus2} & & \mathbb{Z}^{\oplus2} \\
			2 & & & & & & & \\
			1 & & & & & & & \\
			0 & \mathbb{Z} & & & & \mathbb{Z}^{\oplus2} & & \mathbb{Z}^{\oplus2} \\
			\hline
			& 0 & 1 & 2 & 3 & 4 & 5 & 6
		\end{array}
		& \Longrightarrow &
		\begin{array}{c|c}
			6 & \mathbb{Z} \\
			5 & \\
			4 & \mathbb{Z} \\
			3 & \\
			2 & \\
			1 & \\
			0 & \mathbb{Z} \\
			\hline
			&
		\end{array}
	\end{array}
\end{equation}

Upon examining the $E_\infty$ page, we can identify the transgression $d_6:E_6^{0,5}\rightarrow E_6^{6,0}$ which yields $E_7^{0,5}=0$ and $E_7^{6,0}=\mathbb{Z}$. Considering the symmetry of exchanging two $SU$ factors, the generator $x_5\in H^5(SU;\mathbb{Z})$ transgresses to $c_3-c'_3\in H^6(BSU\times BSU;\mathbb{Z})$. Consequently, the normalization can be expressed as
\begin{equation}
    \exp\left(2\pi i\int_{W_5}N_c\frac{x_5}{2}\right),
\end{equation}
which effectively reproduces the anomaly polynomial (\ref{SUAN}) through the transgression. This result is consistent with previous results (\ref{WZWgamma}). Thus, if a manifold has a spin structure, the WZW term reproduces the anomaly and is well-defined.

\section{The four-dimensional $Sp(N_c)$ QCD case}\label{sec4}

As mentioned earlier, $Sp$ gauge theory has been actively studied in recent years in particle phenomenology. In particular, there is a possibility that dark matter can be explained by a strong interaction theory like QCD. We define the $Sp$ WZW term in terms of invertible phases. We observe that the WZW terms are precisely defined on spacetime manifolds with spin structures.

\subsection{Dynamics in UV}

Let us delve into the $Sp(N_c)$ gauge theory, where the fermions are in the fundamental $2N_c$-dimensional representation. This is expressed as
\begin{equation}
    \mathcal{L}=\int_{M_4}\mathrm{d}^4x\left(\tr F_{\mu\nu}F^{\mu\nu}+\bar{\psi}(i\cancel{D}-m)\psi\right).
\end{equation}
Because the fundamental representation of the symplectic group is pseudoreal, it is impossible to tell quarks and antiquarks apart. In this setup, we must have an even number of fermions. If not, a global gauge anomaly arises, rendering the theory mathematically inconsistent. Therefore, going forward, we will focus on the flavor symmetry of $SU(2N_f)$ throughout this paper.

\subsection{Anomalies in UV}

In this section, we discuss quantum anomalies. Initially, you will encounter them as symmetries present in classical theory that do not persist in quantum theory. Here, we focus on anomalies arising from the coupling of fermions to gauge fields. These anomalies exhibit close relationships with various topological aspects of gauge theories and give rise to several interesting phenomena.

There are several approaches to computing anomalies. Let us start by examining how anomalies emerge in a perturbation theory. If we consider a triangular diagram and find that it does not vanish, we can categorize anomalies based on the number of gauge currents and global currents present at the three vertices. 

Firstly, let us consider the scenario where gauge currents are present at all three vertices. This is referred to as a gauge anomaly. It is crucial to recognize that gauge symmetry is not truly a symmetry but rather a redundancy in the theory's description. Consequently, anomalies in gauge symmetry lead to mathematical inconsistencies. It is imperative to ensure the disappearance of all gauge anomalies. The simplest way to avoid gauge anomalies is by making the theory vector-like.

Secondly, let us consider a situation where both gauge and global currents are present at the vertices of a triangle. This is termed a mixing anomaly, for example, known as the chiral anomaly. This phenomenon involves the violation of the conservation law of the axial charge in quantum theory when there is a gauge field coupled to a vector current. Such anomalies in global symmetry are closely related to the topology of gauge theories.

Thirdly, let us consider a scenario where all three vertices exhibit global currents. This implies that the global symmetry has a 't Hooft anomaly. Despite this anomaly, the global symmetry remains intact in quantum theory. However, when coupled to a background gauge field, charge conservation is no longer upheld. Additionally, attempting to couple the current to a dynamic gauge field leads to a gauge anomaly, rendering the theory meaningless. In essence, the 't Hooft anomaly serves as an obstacle to gauging global symmetries. The 't Hooft anomaly is valuable because it offers insights into low-energy dynamics, particularly the spectrum of massless particles, by imposing anomaly matching conditions.

Note that there may also be a global gauge anomaly, which pertains to an anomaly in a gauge symmetry rather than a global symmetry. This type of anomaly is non-perturbative and cannot be observed through perturbative methods like computing triangular diagrams. To see this anomaly, we must examine the global structure of the gauge group. The theory may suffer anomalies under large gauge transformations that cannot be smoothly deformed into trivial gauge transformations. For instance, although all representations of $SU(2)$ are real, suggesting the absence of gauge anomalies, the global gauge anomaly leads to a mathematical inconsistency when an odd number of Weyl fermions is present.

Now, let us consider the general theory of anomalies from a mathematical standpoint. We begin by assuming a spacetime manifold equipped with a spin structure. In this context, a $d$-dimensional anomalous quantum field theory (QFT) possessing symmetry $G$ can be understood as the boundary of a $(d+1)$-dimensional invertible QFT with the same symmetry $G$. When these components are combined, they become gauge invariant. A significant characteristic of anomalies on the boundary is that they can be characterized as the deformation class of the invertible QFT. Specifically, for a symmetry $G$, this deformation class is denoted by $\alpha_G\in\mathrm{Inv}^{d+1}_\mathrm{spin}(BG)$. Here the anomaly polynomial is given by

\begin{equation}
    a(\alpha_G)\in\mathrm{Hom}_\mathbb{Z}(\Omega_{d+2}^\mathrm{spin}(BG),\mathbb{Z}),
\end{equation}
where the map $a$ is in the short exact sequence of the invertible phase as in (\ref{InvSES}):
\begin{equation}
    0\rightarrow\mathrm{Ext}_\mathbb{Z}(\Omega_{d+1}^\mathrm{spin}(BG),\mathbb{Z})\rightarrow\mathrm{Inv}^{d+1}_\mathrm{spin}(BG)\xrightarrow{a}\mathrm{Hom}_\mathbb{Z}(\Omega_{d+2}^\mathrm{spin}(BG),\mathbb{Z})\rightarrow0.
\end{equation}
Note that the anomaly polynomial is the free part of the invertible phase and does not contain the information of the torsion part. In the $\mathbb{Q}$ coefficient, the anomaly polynomial is given by
\begin{equation}
    \mathrm{Hom}_\mathbb{Z}(\Omega_{d+2}^\mathrm{spin}(BG),\mathbb{Z})\otimes\mathbb{Q}=H^{d+2}(BSpin\times BG;\mathbb{Q}),
\end{equation}
which is constructed by a polynomial of spacetime Pontryagin classes and the differential forms on $BG$. In the presence of the spacetime curvature, for a chiral fermion in the representation $V$ of $G$, the anomaly polynomial is given by $(d+2)$-form part of
\begin{equation}
    a\left(\alpha_G(V)\right)=\left[\hat{A}\,\mathrm{ch}(V)\right]_{d+2},
\end{equation}
where $\hat{A}$ means the $\hat{A}$ genus:
\begin{equation}
    \hat{A}=1-\frac{1}{24}p_1+\frac{1}{5760}(7p_1^2-4p_2)+\frac{1}{967680}(-31p_1^3+44p_1p_2-16p_3)+\cdots,
\end{equation}
and $\mathrm{ch}(V)$ is the Charn character in the representation $V$ defined by
\begin{equation}
    \mathrm{ch}(V)=\tr_V\exp\left(\frac{i\mathcal{F}}{2\pi}\right),
\end{equation}
where $\mathcal{F}$ is the gauge curvature. Note that the $i$-th Chern character is represented by $i$-th or lower Chern classes. In particular, when the first Chern class vanishes, the third Chern character is given by one half of the third Chern class.

Let us consider mixed anomalies between gauge symmetries and flavor symmetries. Consider a fermion system with symmetry $G\times F$, where $G$ is anomaly-free and therefore gauged. Furthermore, when mixed anomalies between $G$ and $F$ are absent, the flavor symmetry is given by $F$. The absence of mixing of anomalies is seen in the relation $\alpha_{G\times F}=p^\ast(\alpha_F)$, where $p^\ast$ is the pull-back of the natural surjective group homomorphism $p:G\times F\rightarrow F$. Note that anomalies are elements of $\alpha_{G\times F}\in\mathrm{Inv}^{d+1}_\mathrm{spin}(B(G\times F))$ and $\alpha_F\in\mathrm{Inv}^{d+1}_\mathrm{spin}(BF)$ respectively.

Let us move on to four-dimensional $Sp(N_c)$ QCD with an even number of fermions. We restrict our analysis to the case where $N_c\ge2$ since for $N_c=1$, there exist group isomorphisms such as $Sp(1)\cong SU(2)$. Regarding the flavor symmetry, we examine the scenario of $SU(2N_f)$ with $N_f\ge2$ because, as we will discuss later, for $N_f=1$, the target space $SU(2)/Sp(1)$ associated with the flavor symmetry breaking $SU(2)\rightarrow Sp(1)$ becomes trivial.

Now the fermion multiplets are charged under $Sp(N_c)\times SU(2N_f)$ where $N_c\ge2$ and $N_f\ge2$. Let us examine the anomalies of the $Sp(N_c)$ gauge group. Fortunately, the perturbative gauge anomaly vanishes because any representation of the $Sp$ group is real. This is evident in the vanishing of the free part of the bordism group: $\Omega_6^\mathrm{spin}(BSp(N_c))_\mathrm{free}=0$. As for the non-perturbative gauge anomaly, the associated bordism group is non-trivial: $\Omega_5^\mathrm{spin}(BSp(N_c))=\mathbb{Z}_2$. A global gauge anomaly exists in $Sp$ QCD as well as in $SU(2)$ gauge theory. However, with an even number of fermions, we do not encounter a global gauge anomaly like before. Therefore, our $Sp(N_c)$ gauge symmetry is anomaly free and can be safely gauged.

In order to examine the anomalies of $SU(2N_f)$ flavor symmetry, we need to ensure the non-existence of mixed anomalies. Now, the chiral fermions are in the representations of $\bm{N_c}\otimes\bm{2N_f}$. Then, the anomaly polynomial is given by
\begin{equation}
\label{SpAnomaly}
    2N_c\cdot\mathrm{ch}_3(\bm{2N_f})=N_cc_3,
\end{equation}
where $c_3$ is the third Chern class, and the equality holds since the first Chern class is absent. Note that the fiber dimension of the fundamental representation $\bm{N_c}$ is $2N_c$. It is now clear that the anomaly of $Sp(N_c)\times SU(2N_f)$ is pulled back from the anomaly of $SU(2N_f)$. Therefore, we conclude that there is no mixed anomaly between the gauge symmetry and the flavor symmetry.

We have discussed the anomaly polynomial so far. However, although the anomaly is described in the invertible phase, the anomaly polynomial has no information about its torsion part. In fact, the associated bordism group of the torsion part is non-trivial: $\Omega_5^\mathrm{spin}(B(Sp(N_c)\times SU(2N_f)))=\mathbb{Z}_2$, and it seems that there exists a residual anomaly. When the spacetime dimension is five or less, there is no mixed anomaly between $Sp(N_c)$ and $SU(2N_f)$. Thus, for our four-dimensional model, the generator of the above bordism group is derived from the generator of $\Omega_5^\mathrm{spin}(BSp(N_c))=\mathbb{Z}_2$. Note that $\Omega_5^\mathrm{spin}(BSU(2N_f))$ is trivial, and we ignore it. Now, we have an even number of fermion multiplets, avoiding such a residual anomaly.

\subsection{Dynamics in IR}

Let us move on to the IR physics of $Sp$ QCD. This theory includes gluons and quarks denoted as $\psi_{\alpha ai}$, where $\alpha$ is the spacetime spinor index, $a$ is the color index, and $i$ is the flavor index. When $N_f$ is small enough, the quark condensate occurs, and the physics are described by the non-linear sigma model in the low-energy limit. We assume the quark condensate occurs in the most attractive channel. The flavor symmetry $SU(2N_f)$ is spontaneously broken to $Sp(N_f)$. Note that $Sp(N_f)$ is the largest unbroken symmetry where all quarks acquire masses. The quark condensate is given by
\begin{equation}
    \langle\epsilon^{\alpha\beta}\delta^{ab}\psi_{\alpha ai}\psi_{\beta bj}\rangle=v^3\Sigma_{ij},
\end{equation}
where $v$ is the dynamical scale with mass dimension one, and $\Sigma_{ij}\in\{X\in SU(2N_f)|J_{2N_f}=-X^\top J_{2N_f}X\}\subset SU(2N_f)$, where $J_{2N_f}$ is the $2N_f\times 2N_f$ symplectic matrix. In fact, this is the homogeneous space $SU(2N_f)/Sp(N_f)$ by which the IR physics is described.

\subsection{Anomalies in IR}

In strongly coupled gauge theories, quarks that appear at high energies are often confined to bound states at low energies. As a result, understanding the dynamics at low energies becomes a very difficult problem since the spectrum is complexly changed by quantum effects at low energies. However, the 't Hooft anomaly provides a bit of insight. The 't Hooft anomaly is considered to be invariant under the renormalization group (RG) flow from high energies to low energies. Thus, the anomaly matching condition provides a hint to the strongly coupled region that is difficult to analyze. In the present case, assuming that the Nambu-Goldstone bosons associated with the symmetry breaking $SU(2N_f)\rightarrow Sp(N_f)$ are the only massless spectrum of the low-energy theory, the anomaly matching condition should reproduce the anomaly expressed by (\ref{SpAnomaly}) at high energies. We see that this condition is achieved by the WZW terms.

\subsubsection{Ungauged WZW terms and invertible phases}

Moving forward, there is no need to differentiate the case by $N_f$ in $Sp$ QCD as the number of flavors $N_f$ ($\ge2$) is not relevant to the discussion. This becomes evident from the cohomology ring $H^\ast(BSp(n);\mathbb{Z})=\mathbb{Z}[q_1,q_2,\dots,q_n]$, where $q_i$ represents the $i$-th symplectic Pontryagin class with a degree of $4i$, simplifying the topological structure in the low-degree region.

Consider a configuration $\sigma:M_4\rightarrow SU(2N_f)/Sp(N_f)$ which is extendable to a one dimension higher configuration $\sigma:W_5\rightarrow SU(2N_f)/Sp(N_f)$ satisfying $\partial W_5=M_4$. We can define the WZW term as follows:
\begin{equation}
\label{SpWZW}
    \exp\left(2\pi iN_c\int_{W_5}\Tilde{c}_3\right),
\end{equation}
where $\Tilde{c}_3$ denotes the generator of $H^5(SU(2N_f)/Sp(N_f);\mathbb{Z})\cong\mathbb{Z}$. Later, we will verify that this equation is normalized to correctly reproduce the anomaly at high energies.

Now we can understand the WZW term in terms of the invertible phase. By substituting the required bordism group into the expression represented by (\ref{InvSES}), we obtain the WZW term as an invertible phase.

The related short exact sequence is as follows:
\begin{multline}
    0\rightarrow\mathrm{Ext}_\mathbb{Z}(\Omega_4^\mathrm{spin}(SU(2N_f)/Sp(N_f)),\mathbb{Z})\rightarrow\mathrm{Inv}^4_\mathrm{spin}(SU(2N_f)/Sp(N_f))\\
    \rightarrow\mathrm{Hom}_\mathbb{Z}(\Omega_5^\mathrm{spin}(SU(2N_f)/Sp(N_f)),\mathbb{Z})\rightarrow0.
\end{multline}
Since the torsion part of the invertible phase is trivial: $\mathrm{Ext}_\mathbb{Z}(\Omega_4^\mathrm{spin}(SU(2N_f)/Sp(N_f)),\mathbb{Z})=0$, the invertible phase becomes free: $\mathrm{Inv}^4_\mathrm{spin}(SU(2N_f)/Sp(N_f))\cong\mathrm{Hom}_\mathbb{Z}(\Omega_5^\mathrm{spin}(SU(2N_f)/Sp(N_f)),\mathbb{Z})\cong\mathbb{Z}$. Note that the invertible phase is computed through bordism. Computations using the spin bordism implicitly assume that the spacetime manifold is equipped with a spin structure. This is in contrast to computations using homology, which do not specify a structure on a manifold. It is not trivial that the invertible phase reproduces the WZW term defined by (\ref{SpWZW}). By computing bordism using Adams' method, the relation between the generators of the bordism group and the homology group can be understood (please refer to the Appendix \ref{secC}). Since the $h_0$ tower of the relevant Adams chart starts from the bottom, it is understood that they are equal. Therefore, the invertible phase computed from spin bordism reproduces (\ref{SpWZW}).

\subsubsection{Normalization of ungauged WZW terms}

To define the WZW term accurately, it is necessary to ensure the reproduction of the anomaly obtained by the anomaly matching condition. As discussed earlier, the relation between the WZW terms and the anomaly polynomials is understood from the perspective of transgressions in algebraic topology. Specifically, when the symmetry spontaneously breaks from $SU(2N_f)$ to $Sp(N_f)$, the WZW term $\Gamma\in H^{d+1}(SU(2N_f)/Sp(N_f))$ is transgressed into the anomaly polynomial $\alpha\in H^{d+2}(BSU(2N_f))$ through the transgression associated with the fibration: $SU(2N_f)/Sp(N_f)\rightarrow BSp(N_f)\rightarrow BSU(2N_f)$. When we have all the required cohomology rings, we can compute the transgression $d_n:E_n^{0,n-1}\rightarrow E_n^{n,0}$. For $Sp$ QCD, the LSSS is given by

\begin{equation}
	\begin{array}{ccc}
		E_2^{p,q}=H^p(BSU;H^q(SU/Sp;\mathbb{Z})) & & H^{p+q}(BSp;\mathbb{Z})\\
		\begin{array}{c|ccccccc}
			6 & & & & & & & \\
			5 & \mathbb{Z} & & & & \mathbb{Z} & & \mathbb{Z} \\
			4 & & & & & & & \\
			3 & & & & & & & \\
			2 & & & & & & & \\
			1 & & & & & & & \\
			0 & \mathbb{Z} & & & & \mathbb{Z} & & \mathbb{Z} \\
			\hline
			& 0 & 1 & 2 & 3 & 4 & 5 & 6
		\end{array}
		& \Longrightarrow &
		\begin{array}{c|c}
			6 & \\
			5 & \\
			4 & \mathbb{Z}\\
			3 & \\
			2 & \\
			1 & \\
			0 & \mathbb{Z}\\
			\hline
			&
		\end{array}
	\end{array}
\end{equation}

By considering $E_\infty$ page, the transgression $d_6:E_6^{0,5}\rightarrow E_6^{6,0}$ yields $E_7^{0,5}=0$ and $E_7^{6,0}=0$. This means that the generator $\Tilde{c}_3\in H^5(SU/Sp;\mathbb{Z})$ transgresses to $c_3\in H^6(BSU;\mathbb{Z})$. Thus, the correctly normalized WZW term is given by
\begin{equation}
    \exp\left(2\pi iN_c\int_{W_5}\Tilde{c}_3\right).
\end{equation}
This expression matches the one defined in (\ref{SpWZW}).

\subsubsection{Topological Consistency of gauged WZW terms}

Additional care is needed when defining the gauged WZW terms. They must be defined in such a way that no global anomalies remain. Residual global anomalies are seen in the torsion part of the following commutative diagram of exact sequences: 

\begin{equation}
\begin{tikzcd}
    0\arrow[r] & \mathrm{Ext}_\mathbb{Z}(\Omega_{d+1}^\mathrm{spin}(BH),\mathbb{Z})\arrow[r,"f"] & \mathrm{Inv}^{d+1}_\mathrm{spin}(BH)\arrow[r,"a"] & \mathrm{Hom}_\mathbb{Z}(\Omega_{d+2}^\mathrm{spin}(BH),\mathbb{Z})\arrow[r] & 0\\
    0\arrow[r] & \mathrm{Ext}_\mathbb{Z}(\Omega_{d+1}^\mathrm{spin}(BG),\mathbb{Z})\arrow[r,"f"]\arrow[u,"p^\ast"] & \mathrm{Inv}^{d+1}_\mathrm{spin}(BG)\arrow[r,"a"]\arrow[u,"p^\ast"] & \mathrm{Hom}_\mathbb{Z}(\Omega_{d+2}^\mathrm{spin}(BG),\mathbb{Z})\arrow[r]\arrow[u,"p^\ast"] & 0,
\end{tikzcd}
\end{equation}
where the upward arrows signify pull-backs of $p:BH\rightarrow BG$. We assume that the anomaly polynomial of the IR theory with symmetry $H$ is zero. This means that the anomaly $p^\ast(\alpha_G)\in\mathrm{Inv}^{d+1}_\mathrm{spin}(BH)$ of the IR theory which is the pull-back of the anomaly $\alpha_G\in\mathrm{Inv}^{d+1}_\mathrm{spin}(BG)$ of the UV theory goes to zero $a(p^\ast(\alpha_G))=0$. By the exactness of the upper sequence, there is an element $\tau\in\mathrm{Ext}_\mathbb{Z}(\Omega_{d+1}^\mathrm{spin}(BH),\mathbb{Z})$ where $f(\tau)=p^\ast(\alpha_G)$. A non-zero $\tau$ indicates that the remaining global anomaly in the gauged WZW terms.

For the gauged WZW terms in $Sp$ QCD, the presence of $\mathrm{Ext}_\mathbb{Z}(\Omega_5^\mathrm{spin}(BSp(N_f)),\mathbb{Z})=\mathbb{Z}_2$ suggests the possibility of a residual global anomaly. However, there is no need for concern. It is important to note that the map $p^\ast:H^\ast(BSU(2N_f))\rightarrow H^\ast(BSp(N_f))$ sends the third Chern class to zero, i.e., $p^\ast:c_3\mapsto0$. Additionally, since $\mathrm{Hom}_\mathbb{Z}(\Omega^\mathrm{spin}_6(BSp),\mathbb{Z})=0$, the map $f$ is an isomorphism $\mathrm{Inv}^5_\mathrm{spin}(BSp)\cong\mathrm{Ext}_\mathbb{Z}(\Omega_5^\mathrm{spin}(BSp),\mathbb{Z})$. This implies that $f^{-1}(p^\ast(c_3))=0$, and thus there is no residual global anomaly.

\section{Conclusion}

Considering the recent advancements in $Sp$ gauge theory, we focused on $Sp$ QCD. We studied the WZW terms using bordism theory from the viewpoint of recent developments in invertible phases and anomalies. We complemented the case of $Sp$ QCD, which was lacking in previous studies of $SU$ QCD and $SO$ QCD cases. We explored systems with $Sp$ gauge fields and an even number of fermions, studying the low-energy dynamics when the flavor symmetry spontaneously breaks in the most attractive channel. We confirmed that a well-defined WZW term, capable of reproducing the 't Hooft anomaly, can be described by an invertible phase computed through bordism. We also confirmed that the gauged WZW terms do not create global problems.

\section*{Acknowledgements}
The author is grateful to Hitoshi Murayama for his continuous support in the research. He also provided appropriate advice on this research. Yuji Tachikawa sincerely engaged in discussions and provided useful input. Dan Kondo provided valuable input in proposing and advancing the research. Risshin Okabe complemented my knowledge and provided pertinent guidance in advancing my research. The above four individuals read our pre-finished manuscript and provided very informative and appropriate comments. The author is supported by Forefront Physics and Mathematics Program to Drive Transformation (FoPM), a World-leading Innovative Graduate Study (WINGS) Program, at the University of Tokyo.

\appendix

\section{Cohomology}\label{secA}
This appendix summarizes cohomology rings of Lie groups, classifying spaces, and homogeneous spaces required in this paper. Most of what is written in this appendix is based on \cite{MT91}. The computation of the spin bordism requires cohomology rings of $\mathbb{Z}$ and $\mathbb{Z}_2$ coefficients:
\begin{align}
    H^\ast(BSU(n);\mathbb{Z})&=\mathbb{Z}[c_2,c_3,\ldots,c_n],\label{BSUcohoZ}\\
    H^\ast(BSp(n);\mathbb{Z})&=\mathbb{Z}[q_1,q_2,\ldots,q_n],\label{BSpcohoZ}\\
    H^\ast(SU(n);\mathbb{Z})&=\sideset{}{_\mathbb{Z}}{\bigwedge}[x_3,x_5,\ldots,x_{2n-1}],\label{SUcohoZ}\\
    H^\ast(Sp(n);\mathbb{Z})&=\sideset{}{_\mathbb{Z}}{\bigwedge}[x_3,x_7,\ldots,x_{4n-1}],\label{SpcohoZ}
\end{align}
where $c_i$ is the $i$-th Chern class and the $q_i$ is $i$-th Symplectic Pontryagin class, which has degree $2i$ and $4i$ respectively. The $x_i$ has degree $i$. Note that $\wedge_\mathbb{Z}$ means the exterior algebra with coefficient $\mathbb{Z}$.

In order to obtain cohomology of $SU(2n)/Sp(n)$, we consider the commutative diagram:
\begin{equation}
\label{cohoSUSp_CD1}
\begin{tikzcd}
    H^i(H;\mathbb{Z})\arrow[r,"\bar{\tau}",yshift=0.5ex] & H^{i+1}(BH;\mathbb{Z})\arrow[l,"\bar{\sigma}",yshift=-0.5ex]\\
    H^i(G;\mathbb{Z})\arrow[r,"\bar{\tau}",yshift=0.5ex]\arrow[u,"i^\ast"] & H^{i+1}(BH;\mathbb{Z})\arrow[l,"\bar{\sigma}",yshift=-0.5ex]\arrow[u,equal],
\end{tikzcd}
\end{equation}
where $\bar{\sigma}$ and $\bar{\tau}$ are transgressions induced by the spectral sequences of the following fibrations:
\begin{equation}
\label{cohoSUSp_CD2}
\begin{tikzcd}
    H\arrow[r]\arrow[d,"i"] & EG\arrow[r]\arrow[d,"i_1"] & BH\arrow[d,equal]\\
    G\arrow[r] & G/H\cong EG\times G/H\arrow[r] & BH,
\end{tikzcd}
\end{equation}
where $i$ and $i_1$ are inclusion maps. Note that we consider the case that $G$ is $SU(2n)$ and $H$ is $Sp(n)$. Considering the bottom row of (\ref{cohoSUSp_CD2}), we can obtain the cohomology ring of $SU(2n)/Sp(n)$ by the spectral sequence. By the data of (\ref{BSpcohoZ}) and (\ref{SUcohoZ}), the $E_2^{p,q}$ page $H^p(BSp(n);H^q(SU(2n);\mathbb{Z}))$ is given by
\begin{equation}
\label{SS_SUSp}
    E_2=\mathbb{Z}[q_1,q_2,\ldots,q_n]\otimes\sideset{}{_\mathbb{Z}}{\bigwedge}[\sigma(c_2),\sigma(c_3),\ldots,\sigma(c_{2n})],
\end{equation}
where we use the fact that $x_{2i-1}\in H^{2i-1}(SU;\mathbb{Z})$ is the transgression of $c_i\in H^{2i}(BSU;\mathbb{Z})$ in the spectral sequence of $SU\rightarrow ESU\rightarrow BSU$, that is, $x_{2i-1}=\sigma(c_i)$. Note that the pullback $i^\ast$ in (\ref{cohoSUSp_CD1}) sends the generator of one degree modulo four to zero: $i^\ast(x_{4i+1})=0$, and sends the generator of three degrees modulo four to the generator of $H^\ast(Sp;\mathbb{Z})$: $i^\ast(x_{4i-1})=(-1)^ix_{4i-1}$. Note also that the cohomology of $EG$ (the center of the top row of (\ref{cohoSUSp_CD2})) is trivial in the finite degrees, thus it provides insights into the differentials of the spectral sequence of (\ref{cohoSUSp_CD2}). Consequently the differentials of the spectral sequence (\ref{SS_SUSp}) are given by
\begin{align}
    d_{4i}&:1\otimes\sigma(c_{2i})\mapsto q_i\otimes1,\\
    d_r&:1\otimes\sigma(c_{2i-1})\mapsto0.
\end{align}
Thus, the $E_\infty$ page is computable and consequently the cohomology of $SU(2n)/Sp(n)$ is given by
\begin{equation}
    E_\infty=\sideset{}{_\mathbb{Z}}{\bigwedge}[\sigma(c_3),\sigma(c_5),\ldots,\sigma(c_{2n-1})]=\sideset{}{_\mathbb{Z}}{\bigwedge}[\Tilde{c}_3,\Tilde{c}_5,\ldots,\Tilde{c}_{2n-1}],
\end{equation}
where $\Tilde{c}_i=\sigma(c_i)$ which has degree $2i-1$. Note that the transgressions are commutative with stable cohomology operators, especially the Steenrod squares. This property simplifies the computation of the Adams spectral sequence.

We can obtain $\mathbb{Z}_2$ cohomology by replacing the coefficient ring to $\mathbb{Z}_2$:
\begin{align}
    H^\ast(BSU(n);\mathbb{Z}_2)&=\mathbb{Z}_2[c_2,c_3,\ldots,c_n],\label{BSUcohoZ2}\\
    H^\ast(BSp(n);\mathbb{Z}_2)&=\mathbb{Z}_2[q_1,q_2,\ldots,q_n],\\
    H^\ast(SU(n);\mathbb{Z}_2)&=\sideset{}{_{\mathbb{Z}_2}}{\bigwedge}[x_3,x_5,\ldots,x_{2n-1}],\label{SUcohoZ2}\\
    H^\ast(SU(2n)/Sp(n);\mathbb{Z}_2)&=\sideset{}{_{\mathbb{Z}_2}}{\bigwedge}[\Tilde{c}_3,\Tilde{c}_5,\ldots,\Tilde{c}_{2n-1}].
\end{align}
Note that $\wedge_{\mathbb{Z}_2}$ means the exterior algebra with coefficient $\mathbb{Z}_2$, which is defined to be the polynomial algebra modulo the relations $\forall x\in \wedge_{\mathbb{Z}_2}[\cdots], x^2=0$.

\section{Bordism via Atiyah-Hirzebruch spectral sequence}\label{secB}
In this appendix, spin bodism groups are computed using the Atiyah-Hirzebruch spectral sequence (AHSS) which is a generalization of the Leray-Serre spectral sequence (LSSS) for generalized cohomology. In the case of $BSU$ and $SU$, we referred to \cite{LOT20}. For the fibration $F\rightarrow E\rightarrow B$, the $E^2$ page of AHSS is given by
\begin{equation}
    E^2_{p,q}=H_p(B;\Omega^\mathrm{spin}_q(F)),
\end{equation}
and it converges to $\Omega^\mathrm{spin}_{p+q}(E)$. Taking the trivial fibration $\mathrm{pt}\rightarrow X\rightarrow X$, we can get the spin bordism group from its homology group $H_\ast(X)$:
\begin{equation}
    E^2_{p,q}=H_p(X;\Omega^\mathrm{spin}_q(\mathrm{pt}))\Rightarrow\Omega^\mathrm{spin}_{p+q}(X),
\end{equation}
where the spin bordism of a point is given by
\begin{center}
\begin{tabular}{c|ccccccccc}
    $d$ & 0 & 1 & 2 & 3 & 4 & 5 & 6 & 7 & $\cdots$ \\
    \hline
    $\Omega^\mathrm{spin}_d(\mathrm{pt})$ & $\mathbb{Z}$ & $\mathbb{Z}_2$ & $\mathbb{Z}_2$ & 0 & $\mathbb{Z}$ & 0 & 0 & 0 & $\cdots$
\end{tabular}
\end{center}
Note that because of the splitting of $\Omega^\mathrm{spin}_d(X)=\Omega^\mathrm{spin}_d(\mathrm{\mathrm{pt}})\oplus\Tilde{\Omega}^\mathrm{spin}_d(X)$ the differentials going into the $p=0$ column are all trivial.

\subsection{$BSU(n)$ and $SU(n)$}

In order to specify the WZW terms in $SU$ QCD, the spin bordism group of $BSU(n)$ is required. Here we consider the case of $n\ge3$\cite{LOT20}. In the case of $n=2$, the sixth column of the $E^2$ page disappears and the result changes. The $E^2$ page is given by
\begin{equation}
	\begin{array}{c}
	E^2_{p,q}=H_p(BSU(n);\Omega_q^{\mathrm{spin}})\\
	\begin{array}{c|cccccccc}
		6 &&&&&&\\
		5 &&&&&&\\
		4 & \mathbb{Z} &&&& \mathbb{Z} && \mathbb{Z} &\\
		3 &&&&&& \\
		2 & \mathbb{Z}_2 &&&& \mathbb{Z}_2 && \mathbb{Z}_2 &\\
		1 & \mathbb{Z}_2 &&&& \mathbb{Z}_2 && \mathbb{Z}_2 &\\
		0 & \mathbb{Z} &&&& \mathbb{Z} && \mathbb{Z} &\\
		\hline
		& 0 & 1 & 2 & 3 & 4 & 5 & 6 & 7
	\end{array}
	\end{array}
\end{equation}
The differentials $d_{p,q}^2:E_{p,q}^2\rightarrow E_{p-2,q+1}^2$ in the $E^2$ page are given by the duals of the Steenrod square $Sq^2$ when $q=0,1$\cite{Tei93}. Since the cohomology ring of $BSU(n)$ is given by (\ref{BSUcohoZ}, \ref{BSUcohoZ2}) and the Steenrod square $Sq^2$ leads to the relation $Sp^2c_2=c_3$, the spectral sequence within $p+q\le6$ converges at the $E^3$ page. Thus the spin bordism groups and the reduced ones in the case $n\ge3$ are given by
\begin{equation}
    \Omega^\mathrm{spin}_{d\le6}(BSU(n))=
    \begin{cases}
        \mathbb{Z} & (d=0,6) \\
        \mathbb{Z}_2 & (d=1,2) \\
        0 & (d=3,5) \\
        \mathbb{Z}^{\oplus2} & (d=4)
    \end{cases},
    \Tilde{\Omega}^\mathrm{spin}_{d\le6}(BSU(n))=
    \begin{cases}
        0 & (d=0,1,2,3,5) \\
        \mathbb{Z} & (d=4,6)
    \end{cases}.
\end{equation}

The spin bordism group of $SU(n)$ is also computed in \cite{LOT20}. Since the cohomology ring of $SU(n)$ is given by (\ref{SUcohoZ}, \ref{SUcohoZ2}) and the Steenrod square $Sq^2$ leads to the relation $Sp^2x_3=x_5$\cite{BS53}, the spin bordism groups and the reduced ones in the case $n\ge3$ can be computed as before. The $E^2$ page is given by
\begin{equation}
	\begin{array}{c}
	E^2_{p,q}=H_p(SU(n);\Omega_q^{\mathrm{spin}})\\
	\begin{array}{c|cccccccc}
		5 &&&&&\\
		4 & \mathbb{Z} &&& \mathbb{Z} && \mathbb{Z} &\\
		3 &&&&&\\
		2 & \mathbb{Z}_2 &&& \mathbb{Z}_2 && \mathbb{Z}_2 &\\
		1 & \mathbb{Z}_2 &&& \mathbb{Z}_2 && \mathbb{Z}_2 &\\
		0 & \mathbb{Z} &&& \mathbb{Z} && \mathbb{Z} &\\
		\hline
		& 0 & 1 & 2 & 3 & 4 & 5 & 6
	\end{array}
	\end{array}
\end{equation}
As a result, we obtain the following
\begin{equation}
    \Omega^\mathrm{spin}_{d\le6}(SU(n))=
    \begin{cases}
        \mathbb{Z} & (d=0,3,4,5) \\
        \mathbb{Z}_2 & (d=1,2) \\
        0 & (d=6)
    \end{cases},
    \Tilde{\Omega}^\mathrm{spin}_{d\le6}(SU(n))=
    \begin{cases}
        0 & (d=0,1,2,4,6) \\
        \mathbb{Z} & (d=3,5)
    \end{cases}.
\end{equation}

\subsection{$BSp(n)$}

Let us compute the spin bordism groups of $BSp(n)$. The degrees of the symplectic Pontryagin classes are multiple of four, so $E^2$ page in the range of $p<8$ is independent of $n\ge1$. The $E^2$ page is given by

\begin{equation}
	\begin{array}{c}
        E^2_{p,q}=H_p(BSp(n);\Omega_q^\mathrm{spin})\\
	\begin{array}{c|cccccccc}
		6 &&&&&&&&\\
            5 &&&&&&&&\\
		4 & \mathbb{Z} &&&& \mathbb{Z} &&&\\
		3 &&&&&&&&\\
		2 & \mathbb{Z}_2 &&&& \mathbb{Z}_2 &&&\\
		1 & \mathbb{Z}_2 &&&& \mathbb{Z}_2 &&&\\
		0 & \mathbb{Z} &&&& \mathbb{Z} &&&\\
		\hline
		& 0 & 1 & 2 & 3 & 4 & 5 & 6 & 7
	\end{array}
	\end{array}
\end{equation}

The differential $d^4_{4,1}:E^4_{4,1}\rightarrow E^4_{0,4}$ which arises in $E^4$ page is trivial. So we obtain $E^\infty$ page in the range of $p+q\le6$ which is equal to $E^2$ page. As a result, the bordism groups and the reduced bordism groups of $BSp(n)$ are given by
\begin{equation}
    \Omega^\mathrm{spin}_{d\le6}(BSp(n))=
    \begin{cases}
        \mathbb{Z} & (d=0) \\
        \mathbb{Z}_2 & (d=1,2,5,6) \\
        0 & (d=3) \\
        \mathbb{Z}^{\oplus2} & (d=4)
    \end{cases},
    \Tilde{\Omega}^\mathrm{spin}_{d\le6}(BSp(n))=
    \begin{cases}
        0 & (d=0,1,2,3) \\
        \mathbb{Z} & (d=4) \\
        \mathbb{Z}_2 & (d=5,6)
    \end{cases}.
\end{equation}

\subsection{$SU(2n)/Sp(n)$}

Let us compute the spin bordism groups of $SU(2n)/Sp(n)$. The degrees of generators of $H_\ast(SU(2n)/Sp(n))$ are $0,5,9,\cdots$ in increasing order. So $E^2$ page in the range of $p<8$ is independent of $n\ge2$. The $E^2$ page is given by

\begin{equation}
	\begin{array}{c}
        E^2_{p,q}=H_p(SU(2n)/Sp(n);\Omega_q^\mathrm{spin})\\
	\begin{array}{c|ccccccc}
            5 &&&&&&&\\
		4 & \mathbb{Z} &&&&& \mathbb{Z} &\\
		3 &&&&&&&\\
		2 & \mathbb{Z}_2 &&&&& \mathbb{Z}_2 &\\
		1 & \mathbb{Z}_2 &&&&& \mathbb{Z}_2 &\\
		0 & \mathbb{Z} &&&&& \mathbb{Z} &\\
		\hline
		& 0 & 1 & 2 & 3 & 4 & 5 & 6 
	\end{array}
	\end{array}
\end{equation}

The differential $d^5_{5,0}:E^2_{5,0}\rightarrow E^2_{0,4}$ is trivial. So we obtain $E^\infty$ page in the range of $p+q\le7$ which is equal to $E^2$ page. As a result, the bordism groups and the reduced bordism groups of $SU(2n)/Sp(n)$ for $n\ge2$ are given by
\begin{equation}
    \Omega^\mathrm{spin}_{d\le7}(SU(2n)/Sp(n))=
    \begin{cases}
        \mathbb{Z} & (d=0,4,5) \\
        \mathbb{Z}_2 & (d=1,2,6,7) \\
        0 & (d=3)
    \end{cases},
    \Tilde{\Omega}^\mathrm{spin}_{d\le7}(SU(2n)/Sp(n))=
    \begin{cases}
        0 & (d=0,1,2,3,4) \\
        \mathbb{Z} & (d=5) \\
        \mathbb{Z}_2 & (d=6,7)
    \end{cases}.
\end{equation}

\section{Bordisms via Adams spectral sequence}\label{secC}
There is another way to compute the bordism groups, namely the method of Adams spectral sequence. Adams spectral sequence was invented as a tool to compute stable homotopy groups for spheres. First, instead of computing the set $[Y,X]$, which is the homotopy classes of continuous maps $Y\rightarrow X$, we could consider the induced homomorphisms on the cohomology rings. It is $\mathrm{Hom}_\mathcal{A}(H^\ast(X),H^\ast(Y))$, where $\mathcal{A}$ is mod $p$ Steenrod algebra and the cohomology is taken with the coefficient of $\mathbb{Z}_p$. It is not expected that $\mathrm{Hom}_\mathcal{A}$ has all the information on stable homotopy groups, but given its $s$-th derived functor and grading $t$ by $\mathrm{Ext}_\mathcal{A}^{s,t}$, it is often sufficient. The details of the computation can be found in \cite{LOT20}.

For computing the spin bordism groups, there is a sequence which converges to the 2-completion of the (reduced) spin bordism groups. $E_2$ page is given as
\begin{equation}
    E_2^{s,t}=\mathrm{Ext}_\mathcal{A}^{s,t}(\Tilde{H}^\ast(MSpin\wedge X;\mathbb{Z}_2),\mathbb{Z}_2)\Rightarrow\Tilde{\Omega}^\mathrm{spin}_{t-s}(X)^{\wedge}_2,
\end{equation}
where $\mathcal{A}$ is the mod $2$ Steenrod algebra and $MSpin$ is the Thom spectrum of the universal bundle over $BSpin$. In the range of $t-s\le7$, $E_2$ page is simplified and given by
\begin{equation}
    E_2^{s,t}=\mathrm{Ext}_{\mathcal{A}(1)}^{s,t}(\Tilde{H}^\ast(X;\mathbb{Z}_2),\mathbb{Z}_2),
\end{equation}
where $\mathcal{A}(1)$ is the subslgebra of $\mathcal{A}$ generated by $Sq^1$ and $Sq^2$. By taking the minimal free resolution of $\Tilde{H}^\ast(X;\mathbb{Z}_2)$, we can compute the $E_2$ page.

Let us compute the bordism of $SU(2n)/Sp(n)$. The degrees of the generators of $H^\ast(SU(2n)/Sp(n);\mathbb{Z}_2)$ are five of $\Tilde{c}_3$ and nine of $\Tilde{c}_5$ in increasing order. So when $n\ge2$, the $\mathcal{A}(1)$-module structure of the required cohomology ring $H^\ast(SU(2n)/Sp(n);\mathbb{Z}_2)$ up to degree seven is a single point. In this case, the minimal free resolution is given by
\begin{multline}
    0\leftarrow H^\ast(SU(2n)/Sp(n);\mathbb{Z}_2)\leftarrow \mathcal{A}(1)[5]\leftarrow \mathcal{A}(1)[6]\oplus\mathcal{A}(1)[7]\\
    \leftarrow \mathcal{A}(1)[7]\oplus\mathcal{A}(1)[9]\leftarrow \mathcal{A}(1)[8]\oplus\mathcal{A}(1)[12]\leftarrow\cdots,
\end{multline}
where $[t]$ represents grade-shift. The resulting $E_2$ page is given below where the vertical axis is $s$ and the horizontal axis is $t-s$:

\begin{center}
\DeclareSseqGroup\tower {} {
	\class(0,0)\foreach \i in {1,...,8} {
		\class(0,\i)
		\structline(0,\i-1,-1)(0,\i,-1)
	}
}
\begin{sseqdata}[name=ASS, xscale=0.6, yscale=0.6, Adams grading, classes=fill, xrange={0}{7}, yrange={0}{5}, title={$E_2^{s,t}=\mathrm{Ext}_{\mathcal{A}(1)}^{s,t}(\Tilde{H}^\ast(SU(2n)/Sp(n);\mathbb{Z}_2),\mathbb{Z}_2)$}]
	\tower(5,0)
        \class(6,1)
	\class(7,2)
        \structline(5,0)(6,1)
        \structline(6,1)(7,2)
\end{sseqdata}
\printpage[name=ASS, page=2]
\end{center}
where the vertical and diagonal lines represent the action by $h_0\in\mathrm{Ext}^{1,1}_{\mathcal{A}(1)}(\mathbb{Z}_2,\mathbb{Z}_2)$ and $h_1\in\mathrm{Ext}^{1,2}_{\mathcal{A}(1)}(\mathbb{Z}_2,\mathbb{Z}_2)$ respectively.

There is no non-trivial differential in this $E_2$ page. So one can obtain the 2-completion of the reduced spin bordism groups of $SU(2n)/Sp(n)$:

\begin{center}
\begin{tabular}{c|ccccccccc}
    $d$ & 0 & 1 & 2 & 3 & 4 & 5 & 6 & 7 & $\cdots$ \\
    \hline
    $\Tilde{\Omega}^\mathrm{spin}_d(SU(2n)/Sp(n))$ & 0 & 0 & 0 & 0 & 0 & $\mathbb{Z}_2^\wedge$ & $\mathbb{Z}_2$ & $\mathbb{Z}_2$ & $\cdots$
\end{tabular}
\end{center}

The resulting spin bordism groups via Adams spectral sequence are consistent with the results by AHSS.

\bibliographystyle{utcaps_mod}
\bibliography{ref}

\end{document}